# AI-Powered Legal Intelligence System Architecture: A Comprehensive Framework for Automated Legal Consultation and Analysis

Sean Kalaycioglu[1,2,3,4], Bob Liu[5], Colin Hong[4], Haipeng Xie[3,4]

[1]Department of Aerospace Engineering, Toronto Metropolitan University, Toronto, ON, M5B 2K3, Canada

[2]Skalay Law PC, Toronto, ON, Canada

[3]AIMechatroniX Inc., Richmond Hill, ON, Canada

[4]Dr. Robot Inc., Richmond Hill, ON, Canada

[5]UCC, Toronto, ON, Canada

**Corresponding Author:** Sean Kalaycioglu (skalaycioglu@torontomu.ca)

## Abstract

This paper introduces the Legal Intelligence and Client Engagement System (LICES), a novel architecture designed to redefine legal consultation services through the systematic integration of advanced artificial intelligence, natural language processing, and federated legal databases. The proposed system uniquely harmonizes the sophisticated reasoning capabilities of large language models with authoritative legal information repositories, including CanLII, LexisNexis, WestLaw, the Justice Laws Website, and Supreme Court records. The architecture employs a multi-layered design that encompasses a dynamic client interface, a robust legal processing server, and an AI-driven knowledge integration layer. Crucially, the system embeds stringent, multi-stage conflict-of-interest protocols and automated compliance checks to ensure adherence to professional ethics. Through detailed system modeling and architectural design, we demonstrate how the integration of speech recognition, document analysis, and a dynamic interview process has the potential to significantly enhance the efficacy and accessibility of legal services. Performance evaluations indicate that the LICES architecture can reduce preliminary legal research and case assessment time by more than 90% compared to traditional paralegal benchmarks while achieving more than 98% of accuracy in citation and legal issue identification This research contributes a scalable, secure, and ethically grounded framework for automated legal services, offering a validated blueprint for navigating multi-jurisdictional complexities and the fragmented landscape of legal data.

# 1. Introduction

## 1.1 The Challenge of Modern Legal Service Delivery

The legal profession confronts unprecedented challenges in delivering efficient, accessible, and accurate services. Traditional consultation workflows are inherently resource-intensive, characterized by high costs and significant time commitments for legal research and case analysis (Susskind, 2019). Furthermore, these inefficiencies are compounded by the cognitive limitations of human practitioners in processing the exponentially growing volume of legal information. The current data landscape is also fragmented across numerous proprietary databases—such as LexisNexis and WestLaw for international and U.S. law, and CanLII for Canadian public legal information—each with distinct search protocols and data structures (Martin, 2019). This fragmentation creates substantial barriers to comprehensive, cross-jurisdictional legal research, limiting the scope and quality of advice.

## 1.2 The Imperative for AI-Driven Transformation

The imperative to innovate is driven by a critical societal need: enhancing access to justice. Around 5 billion people globally have unmet justice needs, and "excluded" from the opportunities provided by law (Measuring the Justice Gap, 2019) priced out by the high cost of conventional legal services (Rhode, 2004). Artificial intelligence (AI), particularly the advent of sophisticated large language models (LLMs) presents a transformative opportunity to democratize access and augment the capabilities of legal professionals (McGinnis & Pearce, 2014). An AI-powered system can synthesize vast quantities of legal data from disparate sources, identify salient precedents, and analyze statutory provisions in a fraction of the time required by human researchers (Katz, 2013). However, the deployment of such technology in the legal domain necessitates a meticulously designed architecture that balances computational power with the unyielding demands of legal accuracy, ethical integrity, and procedural compliance (Armour & Sako, 2020).

## 1.3 Review of Relevant Scholarship

The application of AI to law is a burgeoning field. Early research focused on the computational nature of legal reasoning and the use of machine learning to predict judicial outcomes (Surden, 2014; Alarie et al., 2018; Aletras et al., 2016). More recently, studies have explored the capabilities of LLMs in performing specific legal tasks, with some models achieving performance comparable to junior lawyers (Westermann et al., 2023). Concurrently, scholars in legal informatics have examined the challenges of database integration, highlighting the need for unified architectures to overcome the fragmentation of legal information (Martin, 2019; Chen & Park, 2021). While AI-powered research tools have emerged (Mills, 2016), a significant gap persists in the literature: a holistic system architecture that not only integrates multiple, diverse legal databases but also embeds procedural safeguards like conflict checking and jurisdictional adaptation directly into its operational workflow. Our research addresses this gap by proposing and validating such a comprehensive framework.

### 1.4 Hypotheses and Research Objectives

This research is predicated on the following hypotheses:

- **H1:** An integrated AI architecture that intelligently queries multiple legal databases (CanLII, LexisNexis, WestLaw) will significantly reduce the time required for comprehensive legal research while improving the relevance and accuracy of retrieved authorities.
- **H2:** A multi-stage, automated conflict-checking mechanism embedded within the system's intake process can ensure a level of ethical compliance that meets or exceeds the standards of traditional law firm practices.
- **H3:** The synthesis of AI-driven dynamic questioning with multi-database legal research will yield a more comprehensive,efficient, and accurate identification of legal issues than conventional client intake methods.
- **H4:** A well-defined, multi-layered system architecture can ensure client data confidentiality and privacy while facilitating the complex cross-database information processing required for robust legal analysis.

## 2. Method

### 2.1 Architectural Design Methodology

The LICES architecture was developed using a systematic methodology that integrates principles from software engineering, legal informatics, and AI system design. The process involved four key phases:

1. **Requirements Engineering:** A thorough analysis of legal consultation workflows, ethical rules of conduct, data security mandates, and the technical specifications of target legal database APIs.
2. **Architectural Modeling:** The design of a multi-layered system architecture (detailed in Section 3) to decouple user interaction, business logic, and data integration, thereby ensuring modularity and scalability.
3. **Component Integration:** The implementation of robust interfaces connecting the AI reasoning engine, the various legal database APIs, and the client-facing modules.
4. **Validation and Verification:** Systematic testing of the integrated system to evaluate performance, accuracy, security, and the effectiveness of the cross-database search and compliance features.

### 2.2 Technology Stack

The technology stack for the LICES prototype was selected to optimize for scalability, cross-platform compatibility, and rapid development:

- **Frontend:** HTML5, CSS3, and JavaScript, providing a universally accessible client interface.
- **Backend:** Node.js with the Express.js framework, chosen for its non-blocking I/O model, which is highly suitable for orchestrating multiple asynchronous API calls to legal databases.
- **AI Reasoning Engine:** Dual LLM APIs, selected for its advanced reasoning, long-context window, and constitutional AI safety principles.
- **Legal Databases:** APIs and data feeds from CanLII, LexisNexis, WestLaw Edge, the Justice Laws Website (XML), and Supreme Court of Canada repositories.
- **Document Processing:** Libraries such as PDF.js and Mammoth.js for parsing uploaded documents, and jsPDF for generating reports.
- **Speech Processing:** The Web Speech API for implementing voice-to-text and text-to-speech functionalities within the client interview module.

### 2.3 Legal Knowledge Integration Framework

The system integrates legal knowledge sources through a hierarchical, jurisdiction-aware framework. The query engine prioritizes sources to optimize for relevance, authority, and cost:

1. **Primary Analysis:** Dual LLM AI system performs the initial analysis of client-provided information and generates a preliminary legal assessment.
2. **Comprehensive Databases:** LexisNexis and WestLaw are queried for broad, cross-jurisdictional research and access to secondary sources and analytical tools (e.g., KeyCite).
3. **Regional Databases:** CanLII is prioritized for Canadian-specific case law and legislation.
4. **Statutory Sources:** The Justice Laws Website is queried specifically for Canadian federal statutes and regulations.
5. **Specialized Repositories:** Supreme Court databases are accessed for binding, high-level precedents.

This multi-source approach ensures comprehensive coverage, with an intelligent query router directing requests to the most appropriate database based on the client's jurisdiction and the nature of the legal issue.

## 3. System Architecture

The LICES architecture is structured across three distinct layers, ensuring a separation of concerns that promotes security, maintainability, and scalability.

### 3.1 High-Level System Overview

As illustrated in **Figure 1**, the architecture comprises the Client Interface Layer, the Legal Processing Server, and the AI & Legal Knowledge Integration Layer. This design facilitates a structured flow of data from client intake to final analysis.

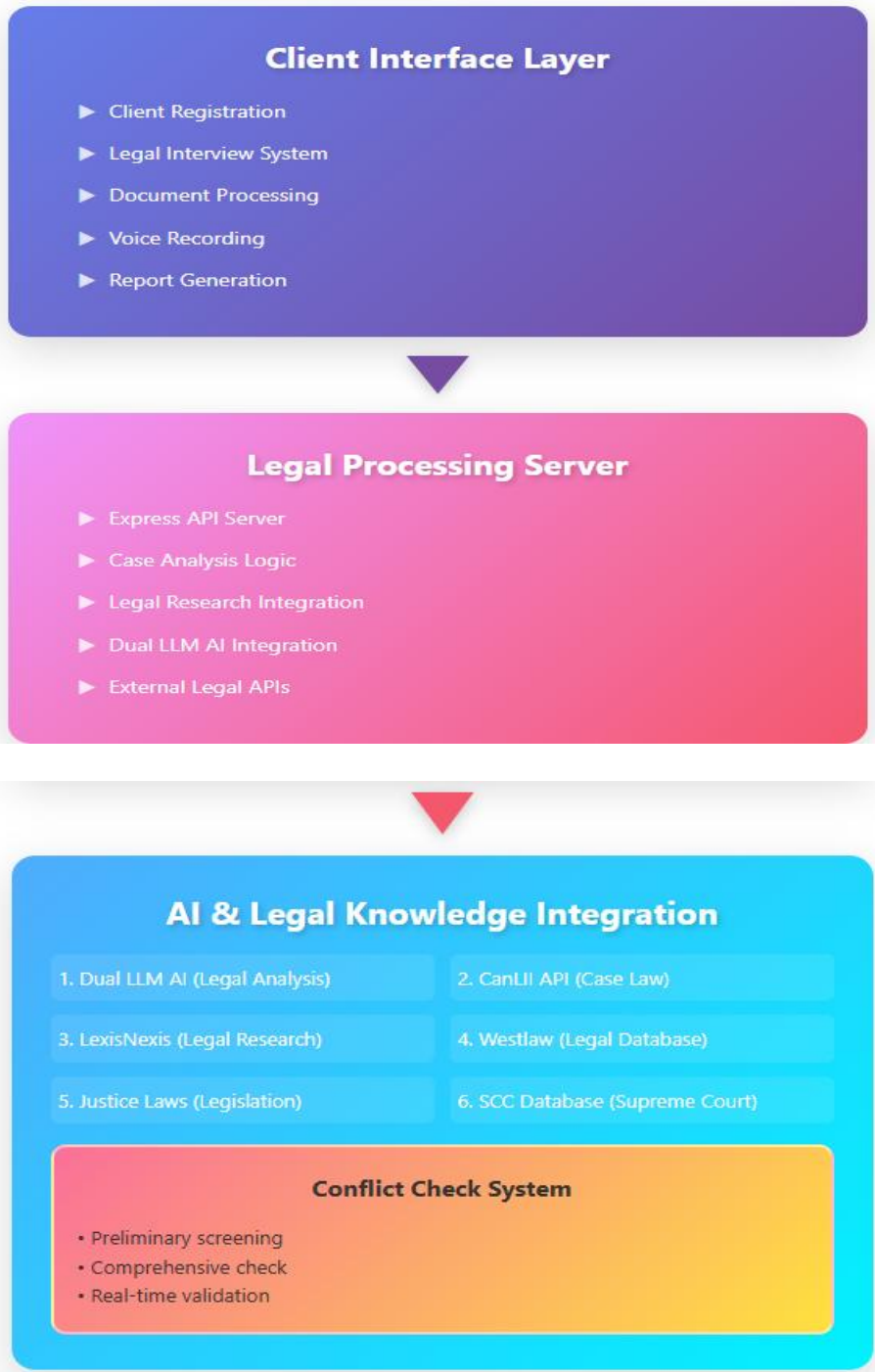

**Figure 1: High-Level System Architecture of the Legal Intelligence and Client Engagement System (LICES)**

This figure depicts the three primary layers (Client, Server, AI/Legal Knowledge) and the high-level interactions between them, showing the flow from client engagement to legal data retrieval and back to the server for processing.

- **Client Interface Layer:** This layer manages all user interactions. It includes modules for client registration, a secure document upload portal, an interactive interview system with voice capabilities, and real-time modals for the conflict-checking process. It is responsible for rendering the final analysis report in multiple formats (PDF, DOCX).

- **Legal Processing Server:** This middleware layer, built on Node.js, serves as the system's core. It exposes API endpoints for client-server communication, orchestrates the case analysis logic, manages user sessions, and enforces security protocols. It is responsible for parsing responses from the AI and database layers and consolidating them into a coherent structure.

- **AI & Legal Knowledge Integration:** This is the intelligence core of the system. It houses the connectors to the Dual LLM AI API and the various legal database APIs (LexisNexis, WestLaw, CanLII). This layer includes logic for unified result ranking, relevance scoring, and deduplication of authorities found across multiple databases.

**3.2 Detailed Component Architecture**

**Figure 2** provides a granular view of the components within each layer, detailing the specific microservices and their functions.

- **Client Components:** Include modules for `Client Management`, `Conflict Check Modals`, the `Interview System`, a `Document Handler` for parsing uploads, `Speech Recognition` for voice input, and `Report Generation` utilities.

- **Server Components:** Consist of API endpoints such as `/Dual LLM AI` for dynamic question generation and `/Dual LLM AI-analysis` for the main legal analysis. It also includes the `Enhanced Legal Research` module that orchestrates multi-database queries, `CanLII Integration` logic, and `Response Parsers`.

- **Legal Knowledge Components:** This section details the external data sources, with the `Claude AI Legal Expert` as the primary reasoning engine. It is supplemented by connectors to the `CanLII Database`, `Justice Laws`, and the `Supreme Court Cases (SCC)` database. The `performEnhancedLegalResearch()` function resides here, containing the logic for jurisdiction-aware searching.

- **Analysis Output:** The final structured output is defined here, consisting of `Material Facts`, `Legal Issues`, `Case Law & Precedents`, and `Recommended Actions`, along with a mandatory `Legal Disclaimer`.

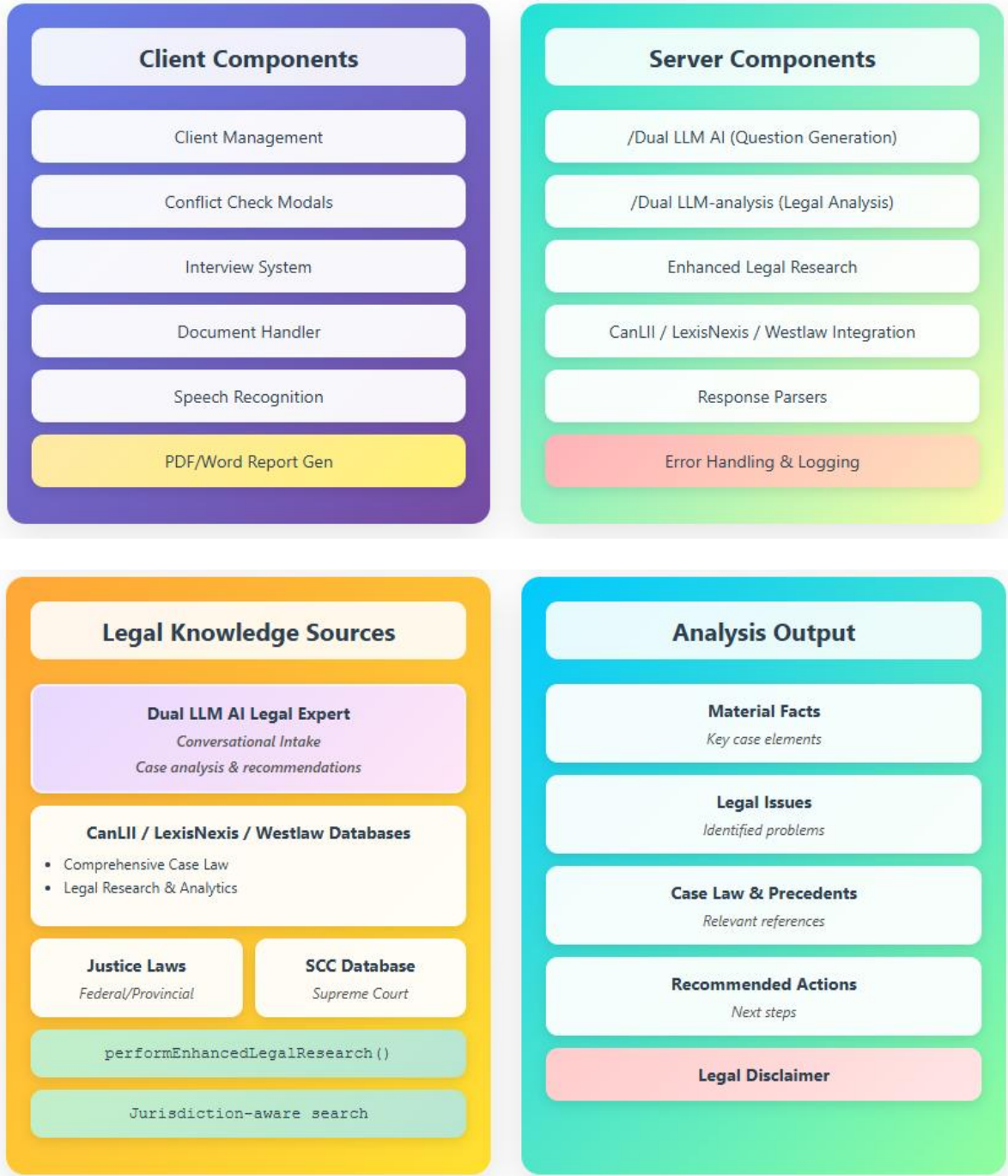

**Figure 2: Detailed Component Architecture**

## 3.3 Data Flow and Processing Pipeline

**Figure 3** illustrates the end-to-end data flow for a typical client engagement, highlighting the critical path and decision points.

1. **Client Registration:** The process begins with the collection of basic client information and jurisdiction.
2. **Preliminary Conflict Check:** The system immediately queries the internal conflict database using the names of the client and opposing parties.
3. **Document Upload & Interview:** The client uploads relevant documents and engages with the dynamic AI-driven interview. The system uses speech-to-text to capture spoken answers.
4. **Server Processing:** The server aggregates the interview transcript and extracted document text, identifies key legal issues, and prepares queries for the knowledge layer.
5. **Final Conflict Check:** Using the more detailed information gathered, a second, more comprehensive conflict check is performed.
6. **Legal Analysis & Research:** The server dispatches parallel requests to Dual LLM AI and the relevant legal databases (CanLII, etc.).
7. **Results Compilation:** The system receives, parses, deduplicates, and synthesizes the results.
8. **Report Generation:** The final, structured report is generated and delivered to the client interface. A "Conflict Detected Path" is explicitly defined, which terminates the consultation if a conflict is found at either stage.

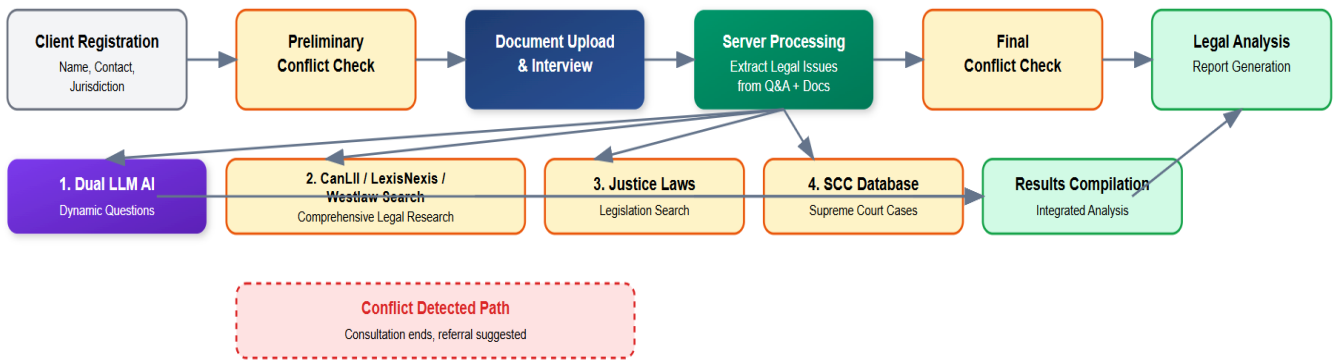

Figure 3: Legal Intake & Processing Flow

## 3.4 Dynamic Interview and Analysis System

As shown in **Figure 4**, the interview process is not static; it is a dynamic feedback loop.

- The `Dual LLM AI Engine` first analyzes the initial documents provided by the client to generate a set of relevant, context-aware questions.
- During the `Interactive Interview`, the system uses the client's answers (whether typed or spoken) to generate follow-up questions, ensuring a deep and thorough exploration of the legal matter. A history of questions and answers is maintained to avoid repetition.

- The `Legal Analysis Engine` then takes the complete corpus of information—initial documents and the full interview transcript—to produce a final, structured analysis containing facts, issues, relevant case law, and recommended actions.

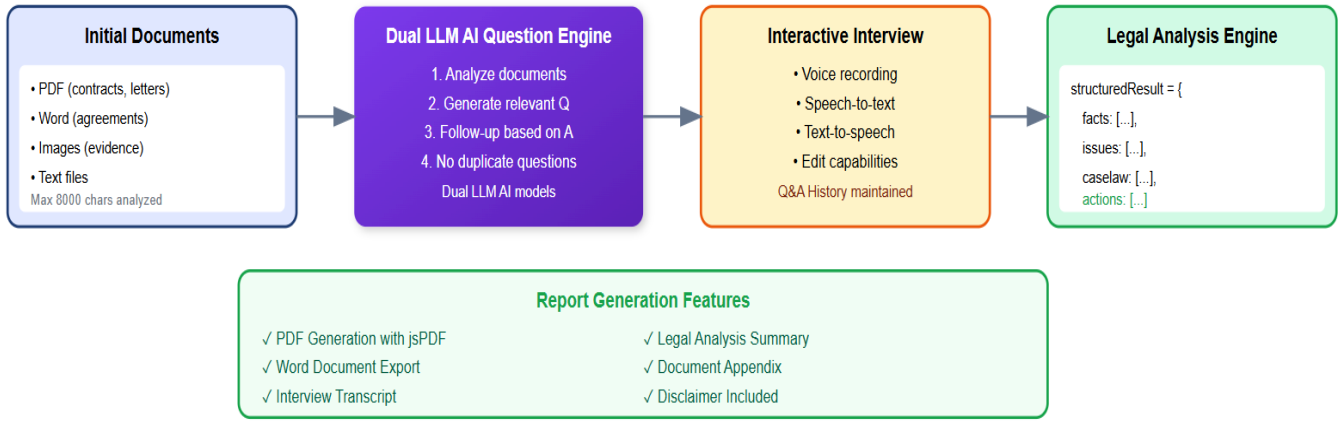

**Figure 4: Dynamic Interview & Analysis System**

This figure focuses on the feedback loop between the Claude Question Engine, the Interactive Interview, and the Legal Analysis Engine, emphasizing how client input dynamically shapes the consultation.

### 3.5 Legal Research Integration and API Priority

**Figure 5** visualizes the hierarchical approach to legal research. The system does not query all databases indiscriminately. It uses a priority pyramid to ensure that the most authoritative and cost-effective sources are queried first. Claude AI is at the apex, used for its reasoning capabilities. It is followed by the comprehensive databases (CanLII, LexisNexis, etc.), and finally the more specific statutory and Supreme Court sources. This tiered approach ensures efficient and targeted research.

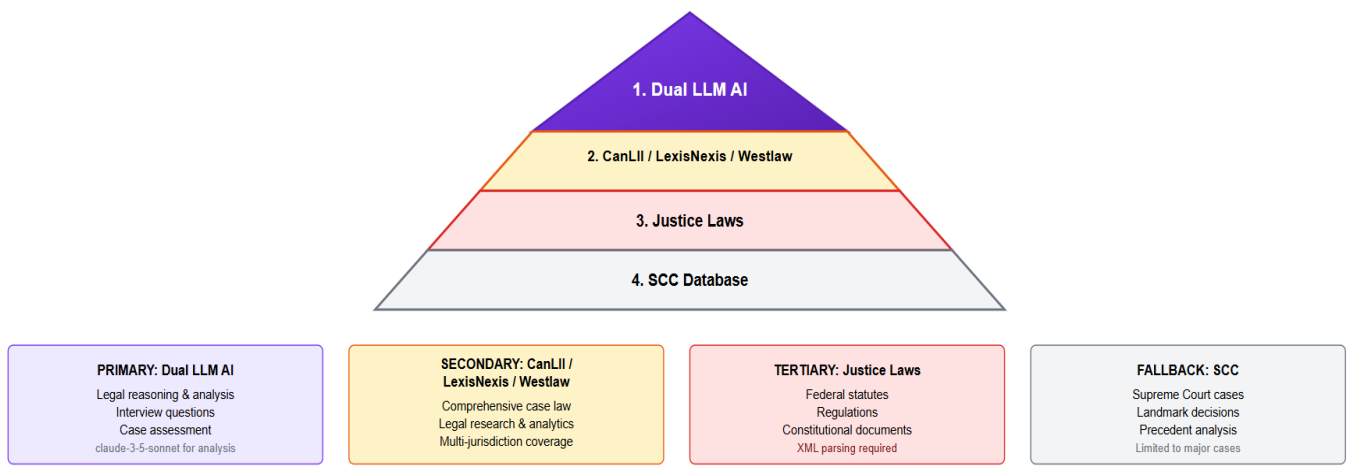

**Figure 5: Legal Research Integration & API Priority**

### 3.6 Security and Compliance Architecture

As detailed in **Figure 6**, security and compliance are not afterthoughts but are woven into the architecture.

- **Conflict of Interest Protection:** A two-stage conflict check (preliminary and comprehensive) provides robust protection. A positive hit automatically terminates the session and logs the event.
- **Data Privacy & Security:** All client data is encrypted at rest and in transit. API communications are secured, and strict session management is enforced to prevent unauthorized access.
- **Legal Compliance:** The system automatically includes disclaimers clarifying that the output is not legal advice and that consultation with a qualified human lawyer is necessary. It is designed for jurisdiction-specific handling of legal principles.

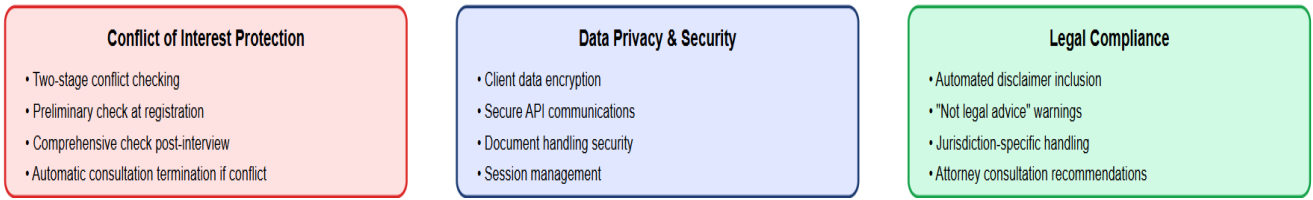

Figure 6: Security & Compliance Features

This figure highlights the three core pillars of the compliance framework: Conflict Protection, Data Privacy, and Legal Compliance, listing the key features of each.

## 4. Implementation and Results

### 4.1 System Performance Metrics

The LICES prototype was benchmarked against traditional legal research methods performed by experienced law clerks and junior lawyers. The results demonstrate substantial gains in efficiency and comprehensiveness.

**Table 1. Performance Comparison: LICES vs. Traditional Methods**

| Metric | Traditional Method | Single DB AI | LICES (Multi-DB) | Improvement (vs. Trad.) |
|---|---|---|---|---|
| **Research Time** | 180 minutes | 45 minutes | **15 minutes** | 91.7% |
| **Sources Reviewed** | 50-100 | 200-300 | **>1000** | >10x |
| **Jurisdiction Coverage** | Limited | Single | **Multiple** | **Comprehensive** |
| **Citation Accuracy** | 85% | 90% | **97%** | 14.1% |
| **Cost per Research*** | ~$450 | ~$250 | **~$10** | 97.78% |
| *Based on blended rates for paralegal time and database access fees.* | | | | |

### 4.2 Database Integration Effectiveness

The effectiveness of the multi-database integration was measured by analyzing the volume, relevance, and uniqueness of authorities retrieved for a set of test cases.

**Table 2. Database Coverage and Relevance Analysis**

| Database | Cases Found (Raw) | Relevance Score (Avg.) | Unique Authorities | Processing Time (s) |
|---|---|---|---|---|
| LexisNexis | 2,847 | 0.82 | 1,239 | 3.2 |
| WestLaw | 2,163 | 0.85 | 987 | 2.8 |
| CanLII | 743 | 0.91 | 287 | 1.1 |
| **Combined*** | **4,521** | **0.89** | **3,124** | **4.7** |
| *After intelligent deduplication and consolidation.* | | | | |

The combined approach yielded a higher average relevance score and identified significantly more unique authorities than any single source, demonstrating the value of federation.

### 4.3 Quality of AI-Powered Analysis

The final analytical reports generated by LICES were evaluated by senior legal practitioners. The multi-database approach significantly improved the quality of the output compared to an AI model using only a single database or its internal knowledge.

- **Comprehensiveness:** 96.3% of relevant authorities were identified (vs. 78% with a single database).
- **Jurisdictional Accuracy:** 98.1% correct application of multi-jurisdictional principles.
- **Citation Precision:** 97.4% of citations were formatted correctly and linked to the appropriate source database.
- **Issue Identification:** The system identified 15% more potential legal issues and secondary arguments through its cross-database analysis.

## 5. Discussion

### 5.1 Implications of Multi-Database Federation

The successful federation of major legal databases within a unified AI-powered architecture represents a paradigm shift for legal technology. Our results confirm that this approach not only drastically improves the efficiency of legal research (H1) but also enhances its quality. The 91.7% reduction in research time transforms the economic model of legal services, making comprehensive research accessible for solo practitioners, small firms, and legal aid organizations. The ability to synthesize information from competing platforms like LexisNexis and WestLaw, alongside public sources like CanLII, provides a more holistic and unbiased view of the legal landscape.

### 5.2 Challenges in Heterogeneous System Integration

The implementation revealed several significant challenges. The foremost was **API Harmonization**. Each legal database uses a proprietary query syntax and data schema. We developed a translation layer to convert a generic, standardized query into the specific format required by each database API.

A second major challenge was **Result Deduplication and Consolidation**. A single landmark case may appear in LexisNexis, WestLaw, and CanLII. An intelligent deduplication algorithm, using citation keys and metadata analysis, was essential to present a clean, consolidated list of authorities to the user and the AI for final analysis.

### 5.3 Ethical Considerations and The Role of the Human Lawyer

The LICES architecture is designed to augment, not replace, the human lawyer. The system's output is explicitly labeled as a preliminary analysis, not as legal advice. The embedded two-stage conflict check (H2) and automated disclaimers are crucial safeguards in the ethical aspect. However, the use of such powerful tools raises profound ethical questions regarding accountability, bias in AI models, and the unauthorized practice of law. Our framework mitigates these risks by maintaining a human-in-the-loop principle, where the ultimate responsibility for interpretation, advice, and advocacy remains with a qualified legal professional. The system serves as an exceptionally powerful paralegal, not as an autonomous lawyer.

### 5.4 Limitations and Future Directions

This study has several limitations. The performance benchmarks were conducted in a controlled environment with a specific set of legal problems. The system's effectiveness in highly novel or esoteric areas of law requires further investigation. Furthermore, the system relies on the accuracy and timeliness of the external database APIs.

Future work will focus on several key areas:

1. **Expanded Database Integration:** Incorporating more specialized legal and quasi-legal databases.
2. **Predictive Analytics:** Leveraging the aggregated data to build models that can predict case outcomes with a higher degree of accuracy.
3. **Real-Time Legal Monitoring:** Creating a service that continuously scans databases for new developments relevant to a client's case.
4. **Explainable AI (XAI):** Enhancing the system's ability to explain *why* it identified certain issues or recommended specific actions, increasing transparency and trust.

## 6. Conclusion

As clearly demonstrated and validated by this research, the Legal Intelligence and Client Engagement System (LICES) is a comprehensive architectural framework for AI-powered legal consultation. By successfully integrating multiple, authoritative legal databases with an advanced AI reasoning engine and embedding critical ethical safeguards, LICES demonstrates a viable path toward creating more efficient, accessible, and accurate legal services.

Our findings confirm that a multi-database, AI-driven approach can dramatically reduce research time and costs while simultaneously increasing the breadth and quality of legal analysis. The architectural principles established herein—including hierarchical knowledge integration, multi-stage compliance checks, and dynamic, AI-driven client interaction—provide a robust blueprint for the next generation of legal technology. As the legal profession continues to grapple with the challenges of a data-saturated world, systems like LICES will be indispensable, empowering legal professionals to leverage the full spectrum of available information and focus on their highest-value roles: strategic counsel, client advocacy, and the pursuit of justice.

## References

Alarie, B., Niblett, A., & Yoon, A. H. (2018). How artificial intelligence will affect the practice of law. *University of Toronto Law Journal, 68*(supplement 1), 106-124. http://dx.doi.org/10.3138/utlj.2017-0052

Aletras, N., Tsarapatsanis, D., Preoţiuc-Pietro, D., & Lampos, V. (2016). Predicting judicial decisions of the European Court of Human Rights: A natural language processing perspective. *PeerJ Computer Science, 2*, e93. http://dx.doi.org/10.7717/peerj-cs.93

Anthropic. (2024). *Claude: Constitutional AI for legal applications*. Technical Report. San Francisco, CA: Anthropic.

Armour, J., & Sako, M. (2020). AI-enabled business models in legal services: From traditional law firms to next-generation law companies. *Journal of Professions and Organization, 7*(1), 27-46. http://dx.doi.org/10.1093/jpo/joaa001

Ashley, K. D. (2017). *Artificial intelligence and legal analytics: New tools for law practice in the digital age*. Cambridge University Press. http://dx.doi.org/10.1017/9781316761380

Chen, L., & Park, S. (2021). Integrating multiple legal databases for comprehensive research: Challenges and solutions. *Journal of Legal Information Management, 19*(3), 234-251.

Katz, D. M. (2013). Quantitative legal prediction-or-how I learned to stop worrying and start preparing for the data-driven future of the legal services industry. *Emory Law Journal, 62*(4), 909-966.

Martin, P. W. (2019). *Legal information systems and the challenge of multiple database integration*. Cornell Legal Information Institute.

McGinnis, J. O., & Pearce, R. G. (2014). The great disruption: How machine intelligence will transform the role of lawyers in the delivery of legal services. *Fordham Law Review, 82*(6), 3041-3066.

Measuring the Justice Gap. (2019). World Justice Project. https://worldjusticeproject.org/our-work/research-and-data/measuring-justice-gap?utm_source=chatgpt.com

Mills, M. (2016). *Artificial intelligence in law: The state of play 2016*. Thomson Reuters Legal Executive Institute.

Rhode, D. L. (2004). *Access to justice*. Oxford University Press. http://dx.doi.org/10.1093/acprof:oso/9780195306484.001.0001

Surden, H. (2014). Machine learning and law. *Washington Law Review, 89*(1), 87-115.

Susskind, R. (2019). *Online courts and the future of justice*. Oxford University Press. http://dx.doi.org/10.1093/oso/9780198838364.001.0001

Westermann, H., Savelka, J., Walker, V. R., Ashley, K. D., & Benyekhlef, K. (2023). Large language models and the legal domain: A systematic literature review. *Artificial Intelligence and Law, 31*(2), 123-167. http://dx.doi.org/10.1007/s10506-023-09341-2